\documentclass[twocolumn]{aastex6}
\pdfoutput=1 
\usepackage[T1]{fontenc}
\usepackage{amsmath,amstext}
\usepackage{apjfonts} 
\usepackage[figure,figure*]{hypcap}
\usepackage{microtype}
\usepackage{caption}
\usepackage{tablefootnote}
\usepackage{etoolbox}
\newcommand{\repeater}{FRB~121102}

\begin{document}
\shortauthors{Eftekhari et al.}
\shorttitle{Potential Analogs of the Repeating FRB 121102}

\title{Wandering Massive Black Holes or Analogs of the First Repeating Fast Radio Burst?}

\author{{T.~Eftekhari}\altaffilmark{1}}
\author{E.~Berger\altaffilmark{1}}
\author{{B.~Margalit}\altaffilmark{2}$^*$}\altaffiliation[*]{NASA Einstein Fellow}
\author{B.~D.~Metzger\altaffilmark{3}}
\author{P.~K.~G.~Williams\altaffilmark{1,4}} 

\altaffiltext{1}{Center for Astrophysics | Harvard \& Smithsonian, Cambridge, MA 02138, USA}
\altaffiltext{2}{Astronomy Department and Theoretical Astrophysics Center, University of California, Berkeley, Berkeley, CA 94720, USA}
\altaffiltext{3}{Department of Physics and Columbia Astrophysics Laboratory, Columbia University, New York, NY 10027, USA}
\altaffiltext{4}{American Astronomical Society, 1667 K St.~NW Ste. 800, Washington, DC 20006}

\begin{abstract}
The discovery of a persistent radio source coincident with the first repeating fast radio burst, \repeater, and offset from the center of its dwarf host galaxy has been used as evidence for a link with young millisecond magnetars born in superluminous supernovae (SLSNe) or long-duration gamma-ray bursts (LGRBs).  A prediction of this scenario is that compact radio sources offset from the centers of dwarf galaxies may serve as signposts for at least some FRBs. Recently, \citet{Reines2020} presented the discovery of 20 such radio sources in nearby ($z\lesssim 0.055$) dwarf galaxies, and argued that these cannot be explained by emission from \ion{H}{2} regions, normal supernova remnants, or normal radio supernovae.  Instead, they attribute the emission to accreting wandering massive black holes. Here, we explore the alternative possibility that these sources are analogs of \repeater.  We compare their properties --- radio luminosities, spectral energy distributions, light curves, ratios of radio-to-optical flux, and spatial offsets --- to \repeater, a few other well-localized FRBs, and potentially related systems, and find that these are all consistent as arising from the same population. We further compare their properties to the magnetar nebula model used to explain \repeater, as well as to theoretical off-axis LGRB light curves, and find overall consistency.  Finally, we find a consistent occurrence rate relative to repeating FRBs and LGRBs.  We outline key follow-up observations to further test these possible connections.
\end{abstract}


\section{Introduction}
\label{sec:intro}

Fast radio bursts (FRBs) are millisecond-duration pulses of coherent radio emission with dispersion measures (DMs) well in excess of the Milky Way contribution, indicative of an extragalactic origin. The sources responsible for producing FRBs remain a topic of debate in large part due to the paucity of well localized events (e.g., \citealt{Eftekhari2017}), as well as to the fact that some FRBs are known to repeat \citep{Chatterjee2017,CHIME2019,Fonseca2020}, while others appear as (so far) single events.

The discovery of the first repeating FRB (\repeater) enabled the first precise localization, leading to the identification of a low metallicity dwarf host galaxy at $z=0.1927$ and an associated persistent radio source \citep{Chatterjee2017,Tendulkar2017,Marcote2017}. The properties of the event and its environment have led to the suggestion that FRBs may be produced by young millisecond magnetars formed in SLSNe or LGRBs \citep{Metzger2017,Murase2016,Piro2016}. This scenario can explain the burst repetitions, the low metallicity host environment (characteristic of SLSNe and LGRBs), and the quiescent radio source (a remnant of the explosion). The recent localizations of three apparently non-repeating FRBs \citep{Bannister2019,Prochaska2019,Ravi2019} and a second repeating FRB \citep{Marcote2020} to more massive galaxies suggest that a sizable fraction of FRB-producing magnetars could possess different formation channels, either some ordinary supernovae or, in the case of an old stellar population, binary neutron star mergers or the accretion-induced collapse (AIC) of a white dwarf \citep{Margalit2019}. 

In the case of the repeating FRB 180916.J0158+65 (hereafter, FRB 180916), radio limits preclude the presence of a persistent radio source to a limit $\sim 300$ times fainter than the \repeater{} persistent radio counterpart. The radio luminosity of an associated FRB nebula is sensitive to the age of the source, however; thus, the lack of a radio counterpart and the low measured rotation measure (RM) (relative to \repeater; \citealt{Michilli2018}) may be indicative of an older system. The discovery of a $\sim 16$-day periodicity in the bursts \citep{Chime2020} has prompted a number of theories, including an FRB-producing magnetar in a tight binary \citep{Ioka2020,Lyutikov2020}, a magnetar undergoing free-precession \citep{Levin2020,Zanazzi2020}, and an ultra-long period magnetar \citep{Paz2020}.

Nevertheless, if some FRBs are produced by millisecond magnetars from SLSNe or LGRBs, then compact radio sources offset from the centers of dwarf galaxies may serve as signposts of these sources. Indeed, a search for such sources \citep{Ofek2017} has led to the identification of the luminous, decades-long radio transient FIRST J141918.9+394036 (hereafter, J1419+3940; \citealt{Law2018}). While a connection to FRBs has not been confirmed for this source, constraints on the source size using the European VLBI Network (EVN) point to an LGRB afterglow origin \citep{Marcote2019}.

A similar search for FRBs and persistent radio emission from the locations of known SLSNe has led to the discovery of a radio source coincident with PTF10hgi nearly a decade after explosion \citep{Eftekhari2019}. The luminosity and age of this source are consistent with central engine powered emission (a magnetar nebula or an off-axis afterglow; \citealt{Eftekhari2019,Law2019}).

Recently, \citet{Reines2020} (hereafter, R20) identified a sample of compact radio sources in nearby ($z\lesssim 0.055$) dwarf galaxies with appreciable offsets from the optical galaxy centers. They rule out \ion{H}{2} regions (i.e., star formation), normal SN remnants, and normal radio SNe as the source of the radio emission, and instead suggest that these sources are powered by accreting massive black holes.  Here we explore the alternative possibility that these sources are analogs of \repeater{} (as well as possibly J1419+4930 and PTF10hgi) using all available information (from R20 and additional data) about the radio luminosities, spectral energy distributions, time evolution, ratio of radio-to-optical emission, and spatial offsets, as well as the expected number of such compact radio sources based on the R20 search. We find consistency in all of these properties, thus warranting consideration that they share a common origin. 

The structure of the paper is as follows. We discuss the sample of radio sources, as well as additional radio observations of each source that we compile from various existing surveys and catalogs in \S\ref{sec:sample}.  In \S\ref{sec:comp} we compare the properties of the sources to \repeater{} and other relevant sources.  We explore theoretical models of off-axis jets and magnetar nebula emission in \S\ref{sec:models}. In \S\ref{sec:rates} we compare the observed occurrence rate of the compact radio sources in dwarf galaxies from the R20 study to the FRB source density and LGRB rate.  We conclude with a discussion of follow-up observations in \S\ref{sec:conclusions}.

\section{Source Sample and Data}
\label{sec:sample}

\subsection{Source Sample}
\label{sec:sourcesample}

R20 observed 111 nearby ($z\lesssim 0.055$) dwarf galaxies with the Karl G.~Jansky Very Large Array (VLA) to search for radio emission at 9 GHz (X-band). The sources were selected by cross-matching Sloan Digital Sky Survey (SDSS) dwarf galaxies ($M_* \lesssim 3\times 10^9 \ M_{\odot}$) with the VLA Faint Images of the Radio Sky at Twenty-centimeters (FIRST) survey \citep{Becker1995}. They detected a total of 48 compact radio sources toward 39 of the 111 targeted galaxies. Among the 48 radio detections, R20 conclude that 20 of the sources are consistent with emission from accreting black holes within the host galaxies.  Here we consider all of these 20 sources, of which 13 are located in confirmed dwarf galaxies (based on redshifts and stellar masses; Sample A in R20) while the remaining 7 do not have reliable redshifts due to insufficient spectra and/or photometry (Sample B in R20); see Table~\ref{tab:vla}. Among the 13 confirmed dwarf galaxies, in all but one case (J0906+5610), emission line diagnostics do not indicate an active galactic nucleus (AGN) origin, as the sources fall within the star forming region of the BPT diagram \citep{Baldwin1981}; as noted in R20, this may not be unexpected for low metallicity galaxies. Finally, among the 7 sources without reliable redshifts, R20 consider J0854$-$0240a and J0854$-$0240b to comprise a candidate dual AGN. 

\subsection{Additional Archival Radio Data}
\label{sec:data}

We collected additional radio data for each of the 20 sources at 3 GHz (S band) from the new VLA Sky Survey (VLASS; \citealt{Lacy2019}). VLASS commenced observations on 2017 September 15 with a typical root-mean-square sensitivity of about 120 $\mu$Jy per epoch and with an angular resolution of about $2.5''$. Quick-look image products are made available within two weeks of observations. We extracted flux densities at the source positions from these images using the \texttt{imtool} function within the \texttt{pwkit}\footnote{https://github.com/pkgw/pwkit.} (version 0.8.20) software package \citep{Williams2017}. 

We also searched the locations of the 20 sources in the NRAO VLA Sky Survey (NVSS; \citealt{Condon1998}) catalog, with observations carried out at 1.4 GHz (L-band) between September 1993 and October 1996 with a completeness limit of about 2.5 mJy. 

Finally, we searched archival VLA data\footnote{https://archive.nrao.edu/cgi-bin/nvas-pos.pl}, as well as NASA's SkyView facility\footnote{https://skyview.gsfc.nasa.gov/current/cgi/query.pl} for additional radio detections across a range of surveys and catalogs (Table~\ref{tab:archival}).  We find a 1.5 GHz Stripe 82 VLA (13B-272) detection for J0106+0046 (2008 September) with a flux density of $1.93\pm 0.10$ mJy; J1200-0341 is detected at 1.4 GHz (AG0644; 2003 February 16) with 7.9 $\pm$ 1.3 mJy; J1136+2643 and J1220+3020 are not detected in archival VLA images covering their locations at 1.5 GHz (AJ0108, AB0506; UT 1984 May 25, 1988 July 12) with $3\sigma$ limits of about 0.7 mJy; and J0134-0741 is detected in the GMRT 150 MHz All-sky Radio Survey with $46.0\pm 8.1$ mJy (2011 November 1).

For each of the 20 sources we determine a minimum age based on a detection in either the NVSS or FIRST surveys. We find that the minimum ages span $8.5 - 26.5$ yr, with a median age of $25.5$ yr. We use this information in \S\ref{sec:models} to investigate the viability of off-axis GRB and magnetar wind nebula models. 

\begin{figure}
\includegraphics[width=\columnwidth]{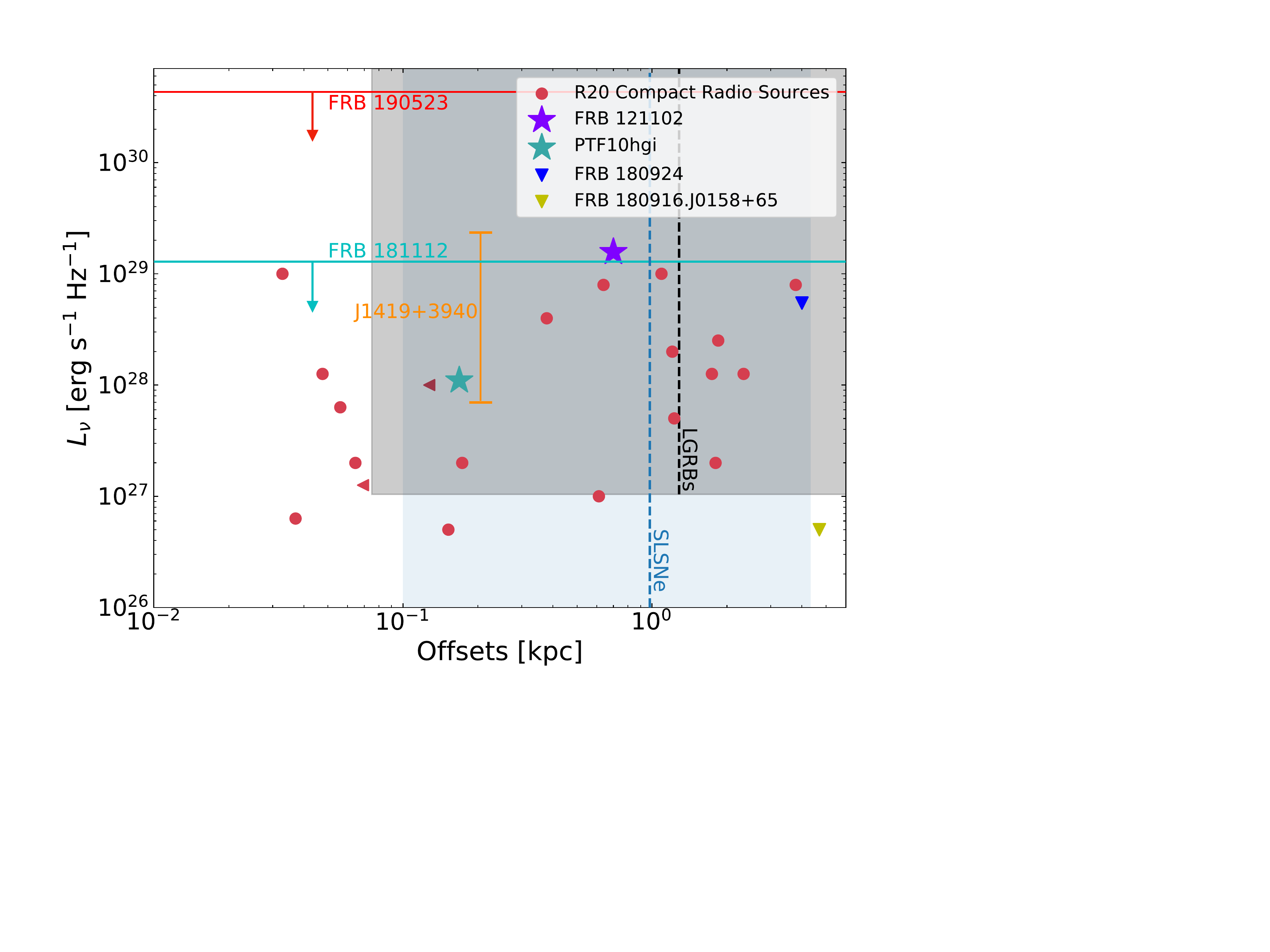}
\caption{Radio spectral luminosities at 9 GHz versus spatial offsets for the 20 compact radio sources from R20 (red circles). Also shown for comparison are the detected radio sources associated with \repeater{} (10 GHz; \citealt{Tendulkar2017}), PTF10ghi (6 GHz; \citealt{Eftekhari2019}), and J1419+3940 (1.4 GHz), for which we plot the range of luminosities that span the 22 year light curve \citep{Law2018}.  We also plot $3\sigma$ upper limits on the radio luminosity of several other localized FRBs, of which two have a well-determined offset (FRBs 180924, 6.5 GHz and 180916, 1.5 GHz) and two (FRBs 181112, 6.5 GHz and 190523, 3 GHz) allow a wide range of offsets. The gray and blue shaded regions correspond to the range of offsets for LGRBs and SLSNe, respectively, with median offsets of $\approx 1.3$ and 1 kpc  \citep{Lunnan2015,Blanchard2016}, although we note that the range of luminosities shown for SLSNe is not representative of the population, as no SLSNe have been detected in the radio to date, with the possible exception of PTF10hgi. For two sources from the R20 sample (J0049$-$0242 and J1226+0815) we plot upper limits on the offsets based on the resolution of the VLA X-band observations ($0.25''$) since R20 report zero offset.}
\label{fig:offsets}
\end{figure}

\section{Comparison to the \repeater{} Persistent Radio Source}
\label{sec:comp}

Here we compare various properties of the 20 compact radio sources from R20 to those of the persistent radio source associated with \repeater{}, the non-detections of persistent radio emission from four additional well-localized FRBs \citep{Bannister2019,Prochaska2019,Ravi2019,Marcote2020}, the late-time radio source associated with the SLSN PTF10hgi \citep{Eftekhari2019}, and the radio transient J1419+3940 \citep{Law2018}.  We focus on the radio luminosities, SEDs, and light curves, the ratio of radio-to-optical flux, and the spatial offsets relative to the host galaxy centers. 

\begin{figure*}
\includegraphics[width=\textwidth]{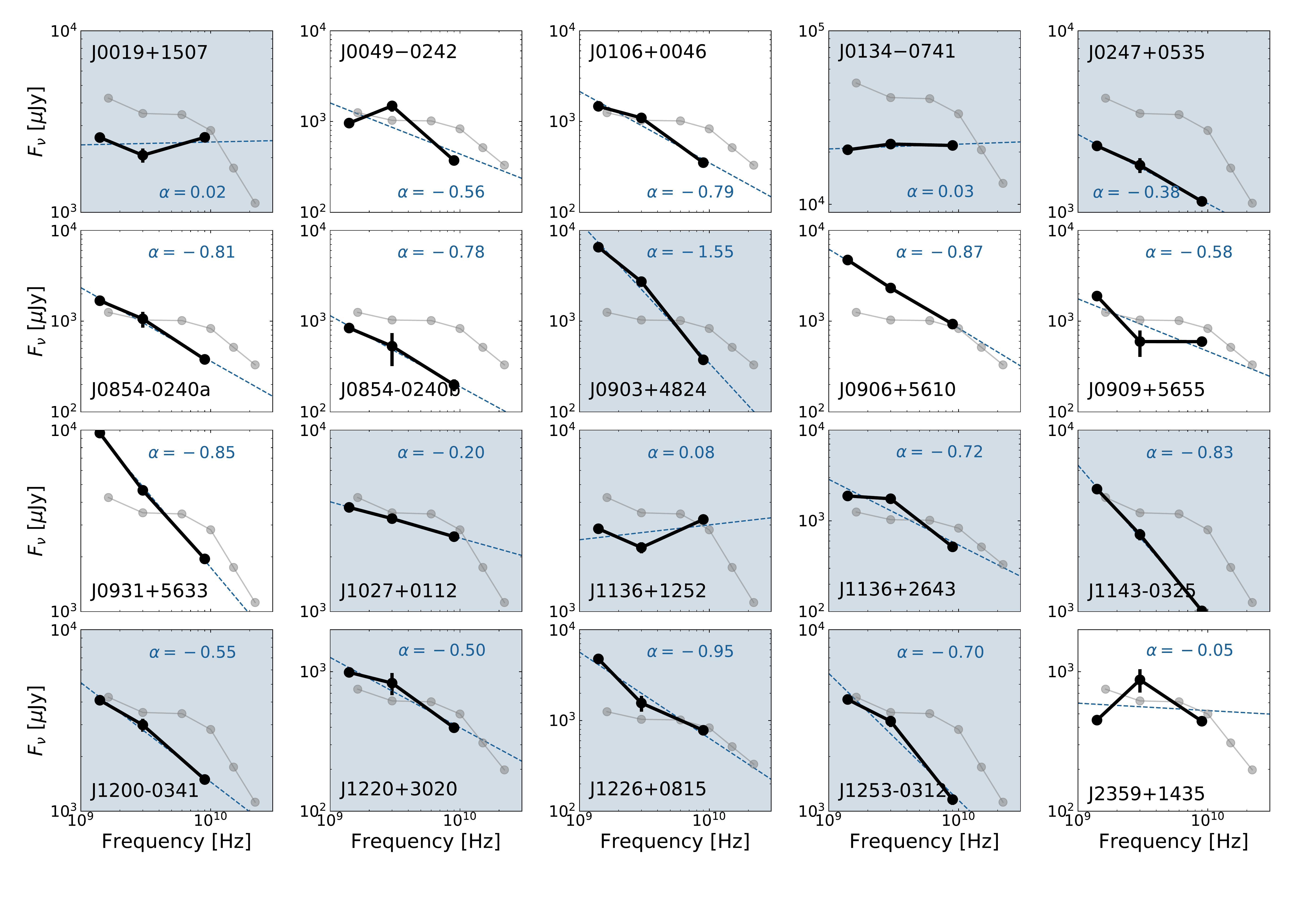}
\caption{Spectral energy distributions (SEDs) for the 20 radio sources from R20 using flux densities from FIRST, VLASS, and R20 at 1.4, 3, and 9 GHz, respectively.  We note that the data points at 3 and 9 GHz were obtained within $\approx 3$ yr, while the FIRST data points span about $5-20$ years prior to the 9 GHz observations. In each panel we also show for comparison the SED of the persistent radio source associated with \repeater{}, with an arbitrary flux scaling (grey points and lines). Dotted blue lines correspond to power-law fits, $F_\nu \propto \nu^{\alpha}$, with the best fit value of $\alpha$ listed in each panel. We highlight individual panels which show strong evidence for time variability at 1.4 GHz based on the lightcurves (see Fig.~\ref{fig:lcs}).}
\label{fig:seds}
\end{figure*}

\subsection{Radio Luminosities}
\label{sec:lums}

In Figure~\ref{fig:offsets} we plot the 9 GHz spectral luminosities of the 20 radio sources from R20 in comparison to detections and limits for well-localized FRBs, PTF10hgi, and J1419+3940.  The R20 sources span a luminosity range of $L_\nu\approx 5\times 10^{26}-10^{29} \ \rm erg \ s^{-1} \ Hz^{-1}$, with a median value of $\approx 1.1 \times 10^{28}\ \rm erg \ s^{-1} \ Hz^{-1}$.  The upper end of the luminosity distribution is comparable to the luminosity of the \repeater{} persistent source, with $1.6\times 10^{29} \ \rm erg \ s^{-1} \ Hz^{-1}$ at 10 GHz. For the 3 additional well-localized (but thus far apparently non-repeating) FRBs, the $3\sigma$ limits on the luminosity of an associated persistent radio source are about $5\times 10^{28}$ (6.5 GHz; FRBs 180924, \citealt{Bannister2019}), $10^{29}$ (6.5 GHz; FRB 181112, \citealt{Prochaska2019}), and $4\times 10^{30}\ \rm erg \ s^{-1} \ Hz^{-1}$ (3 GHz; FRB 190523, \citealt{Ravi2019}). These limits are consistent with the distribution for the R20 sources. 

On the other hand, for the repeating FRB 180916, the $3\sigma$ radio limit is $5 \times 10^{26} \ \rm erg \ s^{-1} \ Hz^{-1}$ (1.5 GHz; \citealt{Marcote2020}), roughly $\sim 300$ times fainter than the \repeater{} persistent source. However, as discussed in \citet{Marcote2020}, this limit is still consistent with a magnetar nebula origin assuming that the system is $\sim 10$ times older than \repeater. This is furthermore consistent with the low observed RM for the source, which is predicted to decay as a function of the source age \citep{Margalit2018b}.

Finally, we find that the range of luminosities is also consistent with the radio emission from PTF10hgi, with $1.1\times 10^{28}\ \rm erg \ s^{-1} \ Hz^{-1}$ (6 GHz; \citealt{Eftekhari2019}), and the radio transient J1419+3940, which spans $\approx 8.5\times 10^{27} - 2\times 10^{29}\ \rm erg \ s^{-1} \ Hz^{-1}$ (1.4 GHz; \citealt{Law2018}) over the time period of about 22 years spanned by its light curve.  

Thus, in terms of radio luminosity we find overall consistency between the R20 sample and our comparison sample of FRBs, PTF10hgi, and J1419+3940.

\subsection{Spectral Energy Distributions}
\label{sec:seds}

In Figure~\ref{fig:seds} we compile the radio SEDs for the 20 sources from R20 using 1.4 GHz flux densities from FIRST, 3 GHz flux densities from VLASS, and 9 GHz flux densities from R20.  We stress that these SEDs are not measured instantaneously, but the observations at 3 and 9 GHz were obtained within $\approx 3$ years, while the FIRST data were obtained about $3-21$ years prior to the date of the 9 GHz observations, with a median time of $\sim 15$ yr prior to the 9 GHz observations. We highlight individual panels where there is strong evidence for time variability at 1.4 GHz based on our light curve analysis in \S\ref{sec:lcs}. Previous studies at $\sim 1-5$ GHz have found that the fraction of variable radio sources at low flux densities ($\sim$ 0.1$-$1 mJy) is $\lesssim 4 \%$, with a variability timescale between minutes and years \citep{deVries2004,Bannister2011,Mooley2013,Mooley2016}. Thus, while in the context of FRB-associated sources the radio emission may vary with time, the time separation between various frequencies is otherwise not problematic given the low occurrence rate for flux variability at the sub-mJy level.

We further note that while J0854$-$0240 is resolved into two distinct radio sources with a separation of $1.7''$ at 9 GHz in the R20 data, the source is blended in the FIRST and VLASS images; we therefore split the flux densities at 1.4 and 3 GHz for these two sources using the ratio of their fluxes at 9 GHz. 

We fit each SED with a single power law, $F_\nu\propto \nu^\alpha$; the resulting values of $\alpha$ are listed in Figure~\ref{fig:seds} and Table~\ref{tab:vla}. We find that most sources exhibit a negative spectral index above 1.4 GHz, with the exception of four sources (J0019+1507, J0134$-$0741, J1136+1252, J2359+1435) that have flat SEDs with $\alpha\approx 0$.  We note, however, that for J1136+1252 the 1.4 GHz data were obtained about 15 years before the VLASS and R20 data. The median spectral index for the sample is $\approx -0.7$, consistent with typical values for mJy-level radio sources \citep{Condon2002}, as well as the expectation for LGRB afterglows at late times \citep{Granot2014}.

We also compare the SEDs to that of the \repeater{} persistent source, which is roughly flat at $1-6$ GHz, and turns over to a power law with $\alpha\approx -0.9$ up to about 22 GHz (Figure~\ref{fig:seds}). This SED is overall consistent with the range of SEDs we find for the R20 sources. Finally, we note that in the case of J1419+3940, the spectral index evolves from $\alpha > 0.6$ at peak to $\alpha\sim -0.6$ at late times, consistent with the median observed spectral index of the R20 sources.

\begin{figure*}
\includegraphics[width=\textwidth]{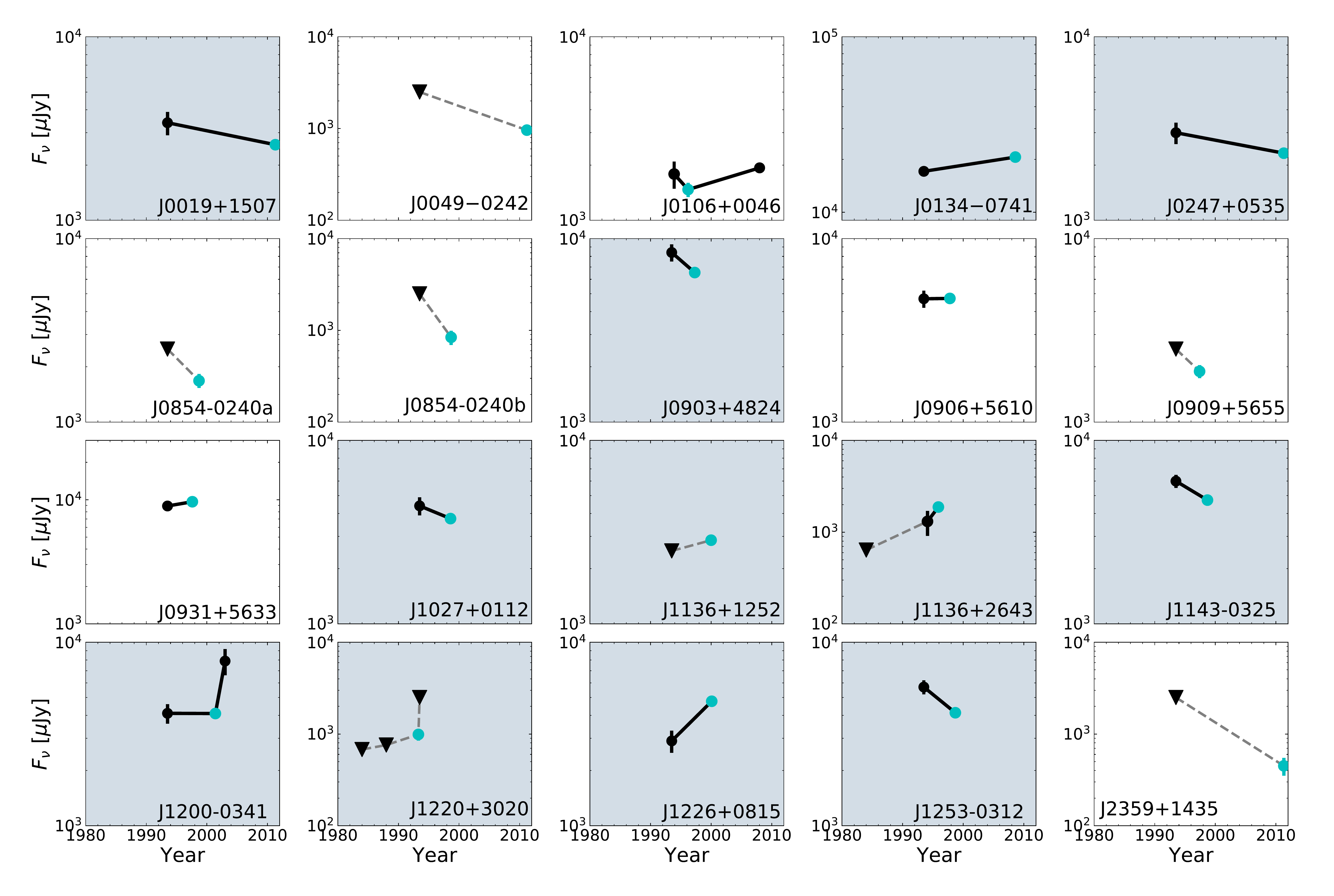}
\caption{Radio light curves at 1.4 GHz from FIRST (cyan), NVSS (black), and a few additional archival data (black; see \S\ref{sec:sourcesample}). Sources with strong evidence for time variability are individually flagged. Triangles and dashed lines mark $3\sigma$ upper limits. We note that the error bars for the NVSS data points refer to the reported peak flux error from the NVSS catalog and thus may be an underestimate of the error on the total integrated flux (plotted); however, given the compact nature of the sources, we expect this difference to be negligible.}
\label{fig:lcs}
\end{figure*}

\subsection{Light Curves}
\label{sec:lcs}

We compile light curves at 1.4 GHz for the R20 sources using data from FIRST and NVSS (Figure~\ref{fig:lcs}). For four of the sources (J0106+0046, J1136+2643, J1200$-$0341, J1220+3020) we include additional archival VLA observations (\S\ref{sec:sample}). We note that 3 of the sources (J0049$-$0242, J0854$-$0240a, J0854$-$0240b) are nominally detected in NVSS, but we cannot derive accurate flux densities due to confusion with nearby bright sources. For 13 of the 20 sources we find at least 2 detections, where 2 of the sources have 3 detections. For 2 sources we have early upper limits that provide useful constraints on flux evolution (i.e., rising flux), and for the remaining 4 sources we have non-constraining early upper limits that do not provide useful information about time evolution.

Among the 13 sources with more than a single epoch detection, the light curves exhibit a range of time evolution. We find that 4 of the light curves show evidence of rising, 6 exhibit a decline, and 3 are essentially flat. An additional 2 sources (J1136+2643 and J1220+3020) have useful upper limits at early times which indicate rising emission. We highlight the sources with strong evidence for time variability in Figure~\ref{fig:lcs}. The 4 rising sources rise by about $20-90\%$ on time baselines spanning $2-15$ years. The 6 declining sources decline by about $20-40\%$ over time baselines of about $4-18$ years.  Finally, the 3 sources with no clear evidence for time variability have baselines of about $5 - 14$ years. We note that the difference in resolution between the NVSS and FIRST surveys ($45''$ and $5''$, respectively) may lead to flux variability at the level of $\sim 30\%$ for extended sources as some of the emission may be resolved out in the FIRST images. From \citet{Reines2020}, 7 of the 20 sources are not designated as point sources as determined by the CASA IMFIT task and based on the resolution of the VLA 9 GHz observations ($0.25''$). We identify 2 of these (J0903+4824 and J1253-0312) as showing evidence for time variability at the level of $\lesssim 30\%$. J1136+2643 and J1252-0312 are also not consistent with being point sources, but both exhibit flux variability at the level of $\gtrsim 30\%$. The 3 remaining sources that are not consistent with being point sources (J0049-0242, J0854-0240a, and J2359+1435) have only single epoch detections and thus we cannot assess their variability.

Studies of the long-term radio variability of AGN suggest that flares typically last $\sim 2-3$ yr \citep{Hovatta2008,Nieppola2009}. Given the sparse temporal coverage of each source, we cannot rule out evolution due to AGN variability. The persistent source associated with \repeater{} does not show obvious evidence for flux evolution on a timescale of $\approx 1$ yr, although variability at the $10\%$ level is seen on day timescales and at the $50\%$ level on a timescale of a week, in both cases likely due to interstellar scintillation \citep{Chatterjee2017}. Thus, the flux evolution that we observe for the R20 sources is at present consistent with the data for \repeater.

\subsection{Spatial Offsets}
\label{sec:offsets}

R20 find that roughly half of the sources in their sample are offset from the optical host galaxy centers by $\gtrsim 0.1$ kpc. They attribute the offsets to velocity kicks following the merger or interaction between two black holes. This is possibly supported by the apparent correlation with how centrally concentrated the galaxies appear to be. In contrast, using precision astrometry from GAIA DR2 to constrain the photometric centers of galaxies, \citet{Shen2019} find that the majority ($99\%$) of AGN at $z\approx 0.3-0.8$ are well-centered to $\lesssim 1$ kpc. As pointed out by \citet{Shen2019}, this difference may arise from the fact that the R20 galaxies are less massive and therefore have shallower gravitational potentials. Furthermore, we note that the optical center is poorly defined in some cases where the host galaxies possess irregular morphologies. 

In Figure~\ref{fig:offsets} we show the offsets of the 20 R20 radio sources, the FRBs, PTF10ghi, and J1419+3940. We also plot the offset distribution of LGRBs \citep{Blanchard2016} and SLSNe \citep{Lunnan2015}. We find that for the R20 sample the overall range of offsets, and the median of $\approx 0.5$ kpc, are comparable to those of \repeater{} (0.7 kpc; \citealt{Marcote2017}), PTF10hgi (0.2 kpc; \citealt{Eftekhari2019}), and J1419+3940 (0.2 kpc; \citealt{Law2018}); the offset for FRB 180924 (for which no persistent radio counterpart has yet been detected) is comparable to the upper end of the distribution for the R20 sources. For FRBs 181112 and 190523 only upper limits on the offset are available of about $27$ and $40$ kpc, respectively, based on the size of the localization regions (a few arcseconds; \citealt{Prochaska2019,Ravi2019}). On the other hand, for the repeating FRB 180916, the offset of $4.7$ kpc is larger than the range of observed offsets for the R20 sources. The position of the FRB coincides with a distinct region of star formation, possibly indicative of an interaction with a dwarf satellite companion. In this context, the large observed offset would be a consequence of the interaction. Thus, in terms of spatial locations within their host galaxies, the R20 sources are consistent with the FRBs, PTF10hgi, and J1419+3940. However, we note that the host galaxies of FRBs 180924, 180916, 181112, and 190523 are generally more massive than the \repeater{} host and the R20 sources, and thus we might expect a different offset distribution for these sources. 

Finally, we note that the distribution of offsets for the R20 sources is largely consistent with the observed distributions for LGRBs and SLSNe, for which the median offsets are $1.3$ and $1$ kpc, repsectively \citep{Lunnan2015,Blanchard2016}. 

\begin{figure}
\includegraphics[width=\columnwidth]{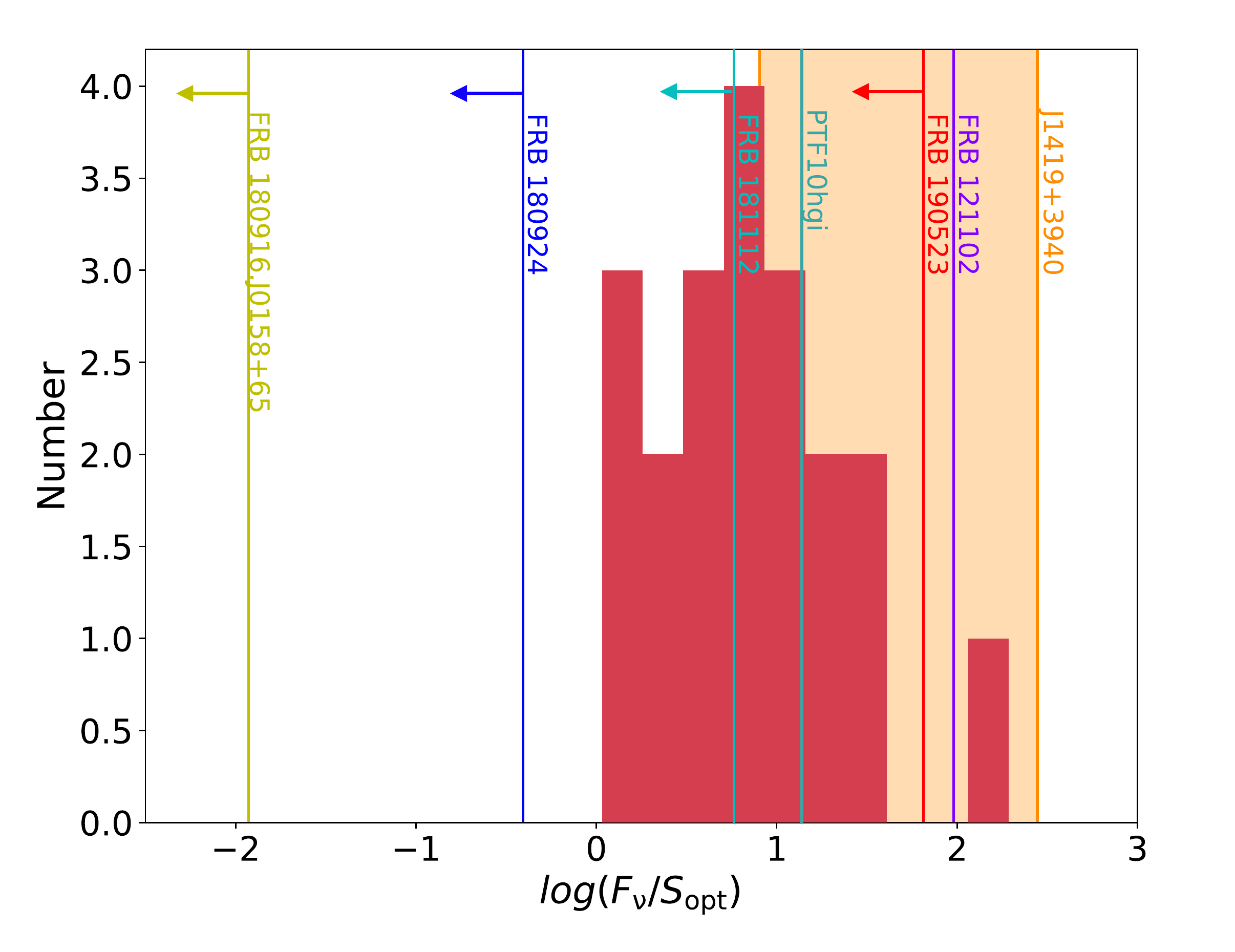}
\caption{Histogram of radio-to-optical flux density ratios defined as $R = \rm{log}(F_{\rm 1.4 GHz}/S_{r})$ for the R20 sources (red) using the 1.4 GHz flux densities from FIRST and SDSS $r$-band magnitudes for the host galaxies.  We also show the values for \repeater{} \citep{Tendulkar2017} and PTF10hgi \citep{Eftekhari2019}. For J1419+2940, we plot a band corresponding to the range of radio fluxes observed over the 22 year duration of the light curve \citep{Law2018}. For FRBs 180916, 180924, 181112, and 190523, we plot upper limits corresponding to their radio non-detections.}
\label{fig:fluxratio}
\end{figure}

\subsection{Ratio of Radio-to-Optical Flux}
\label{sec:sfrs}

In Figure~\ref{fig:fluxratio}, we plot the ratio of radio-to-optical flux densities for the R20 sources.  We use a slightly modified 
version of this ratio than \citet{Padovani2009}, replacing the $V$-band flux density with the SDSS $r$-band, $R\equiv \rm{log}(F_{\rm 1.4 GHz}/S_{r})$. For the radio fluxes we use the 1.4 GHz data from FIRST, given that some of the sources are below the detection limit of NVSS. In each case, we correct the optical flux densities for Galactic extinction \citep{Schlafly11,Barbary16}. 

In general, we find that the combination of low-luminosity dwarf host galaxies and bright radio sources leads to $R\approx 0-2.2$. In contrast, star forming galaxies have $R\lesssim 1.4$ \citep{Machalski1999,Padovani2009}, limited by the radio power of relativistic electrons in the interstellar medium. As argued already by R20, the inferred radio star formation rates (SFRs) significantly exceed the optical SFRs indicating that the radio emission is not due to star formation activity.

For \repeater, a bright persistent radio source coupled with a low-luminosity dwarf host leads to $R\approx 2$ \citep{Chatterjee2017}. Similarly, for J1419+3940, the range of radio luminosities observed over the 22 year duration of the light curve leads to $R\approx 0.9 - 2.4$ \citep{Law2018} and for PTF10hgi we find $R \approx 1.1$ \citep{Eftekhari2019}. The lower radio-to-optical flux ratios observed for some of the R20 sources may therefore reflect the older age of these systems relative to \repeater{}. We note that for the remaining well-localized events (FRBs 180924, 181112, and 190523), limits on radio counterparts lead to limits on the radio-to-optical flux ratio that are consistent with the R20 sources for FRBs 181112 and 190523, while the limits for FRBs 180916 and 180924 lie below the distribution. In particular, the deep radio limit for FRB 180916 leads to a ratio of $-1.9$, well below the distribution for the R20 sources. However, as noted earlier, the coincidence of the FRB position with a peculiar region of star formation points to a possible merger with a dwarf satellite companion. In this scenario, assuming the FRB originates from the dwarf companion, the radio-to-optical flux ratio would shift to higher values. 

Thus, the distribution of radio-to-optical flux density ratios that we observe for the R20 sources is consistent with \repeater{}, J1419+3940, PTF10hgi, and 2 of the 4 additional well-localized FRBs. If FRBs are produced via multiple formation channels (e.g., magnetars born in binary neutron star mergers in addition to SLSNe and LGRBs), or represent magnetars across a range of evolutionary stages \citep{Paz2020}, then the distribution of radio-to-optical flux densities may reflect this.

\section{Models}
\label{sec:models}

\subsection{Off-Axis Jets from LGRBs}
\label{sec:lgrbs}

\begin{figure}
\includegraphics[width=1.1\columnwidth]{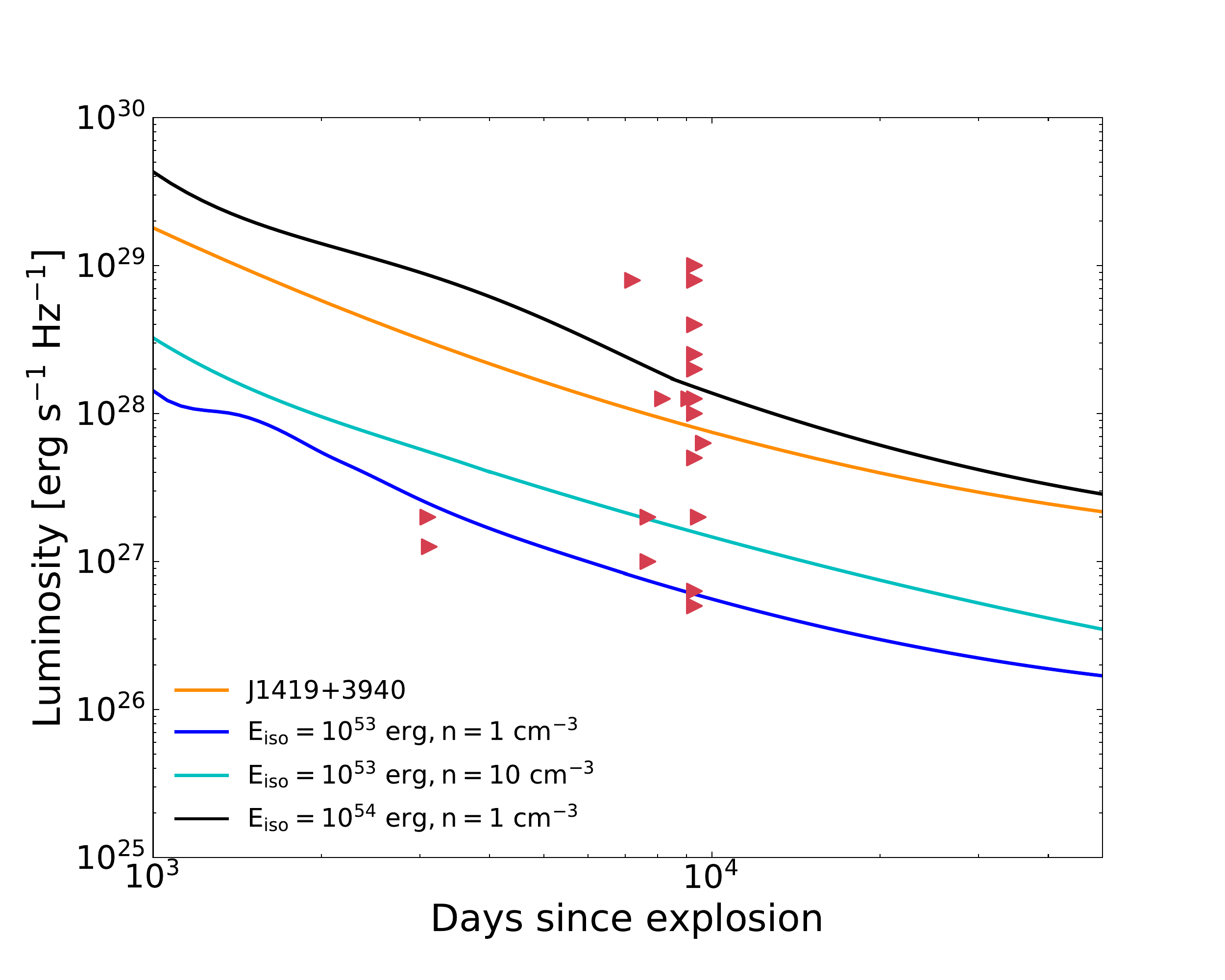}
\caption{Representative off-axis LGRB afterglow models at 9 GHz for $E_{\rm iso}=10^{53}$ and $10^{54}$ erg, $n=1$ and $10$ cm$^{-3}$, and $\theta_{\rm obs} = 60^\circ$. In each case, we assume $\epsilon_e = 0.1$, $\epsilon_B =0.01$, and $p=2.2$. Also shown is the model 9 GHz light curve of J1419+3940 using the best-fit parameters from \citet{Law2018}.}
\label{fig:lgrbs}
\end{figure}

Here we explore the possibility that the R20 sources are due to initially off-axis LGRB jets that have decelerated and spread into our line of sight at late times \citep{Rhoads1997,Sari1999}.  We use the minimum age of each source (based on detections in FIRST or NVSS) and compare the measured luminosities to the expectation for off-axis jet light curves; see Figure~\ref{fig:lgrbs}. We generate radio light curves using the 2D relativistic hydrodynamical code \texttt{BOXFIT v2} \citep{vanEerten2012} for isotropic jet energies $E_{\rm iso} = 10^{53}$ and $10^{54}$ erg, circumburst densities $n=1$ and $10$ cm$^{-3}$, and assuming an initial jet opening angle of $\theta_j = 10^\circ$. We further impose an observer viewing angle of $\theta_{\rm obs}=60^\circ$ and microphysical parameters of $\epsilon_e=0.1$, $\epsilon_B=0.01$, and $p=2.2$, as is typically observed in LGRBs (e.g., \citealt{Curran2010,Laskar2013,Wang2015,Laskar2016,Alexander2017}). We find that these representative models generate radio luminosities comparable to at least some of the R20 sources at the inferred minimum timescales.

We also compare the luminosities and timescales of the R20 sources to the inferred afterglow model for J1419+3940, which \citet{Law2018} fit with $E_{\rm iso}=10^{53}$ erg, $n=10$ cm$^{-3}$, $\theta_{\rm obs} = 34^\circ$, $p=2.2$, $\epsilon_e = 0.1$, and  $\epsilon_B = 0.025$.  This model predicts a 9 GHz luminosity of $\sim 10^{28} \ \rm erg \ s^{-1} \ Hz^{-1}$ at $\sim 25$ years (the median minimum age of the R20 sources), consistent with the median luminosity of the R20 sources. 

We therefore conclude that off-axis LGRBs provide a viable explanation for at least some of the R20 sources given the luminosities and minimum age constraints. Continued monitoring to constrain the light curve evolution will better test this scenario.

\begin{figure*}
\includegraphics[width=\textwidth]{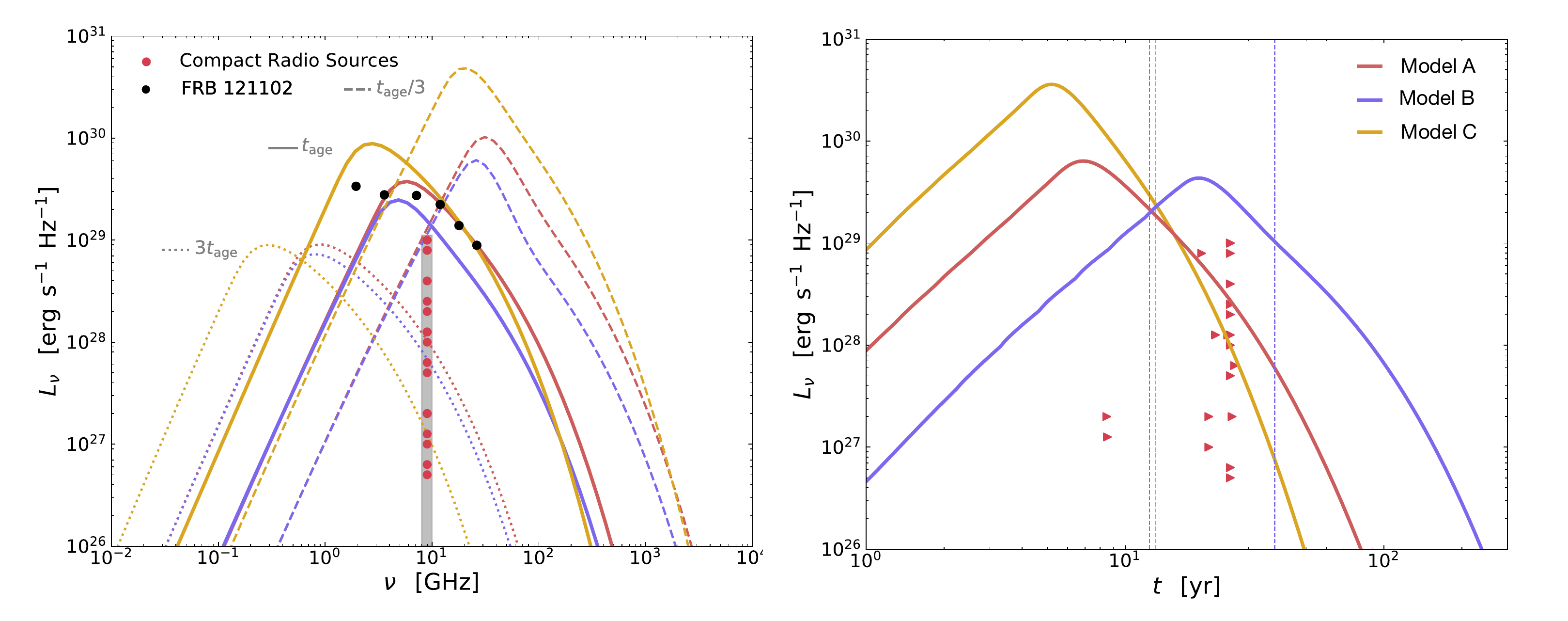}
\caption{Left: Spectral energy distributions for the magnetized electron-ion nebula model invoked to explain the \repeater{} persistent radio source \citep{Margalit2018b} with the compact radio sources at 9 GHz (red points) plotted for comparison. The red, blue, and yellow curves correspond to models A, B, and C from \citet{Margalit2018b}, and the solid, dashed, and dotted curves depict the SED at the observed epoch $t_{\rm age}$, $t_{\rm age}/3$, and $3t_{\rm age}$, respectively. Right: Light curves at 9 GHz for each model as before. Vertical dashed lines correspond to the presumed age of the \repeater{} persistent source for each model. A minimum age constraint for each source is based on a detection in either NVSS or FIRST. The observed luminosities and range of timescales for the compact radio sources are consistent with the models for the \repeater{} persistent radio source.}
\label{fig:frbmodels}
\end{figure*}

\subsection{Magnetar Nebula}
\label{sec:magnetar}

We next analyze the R20 sources in the context of a magnetar nebula model invoked to explain the persistent radio source of \repeater{}, and potentially PTF10hgi and J1419+3940. \citet{Margalit2018b} find that the properties of \repeater{} can be explained in the context of a magnetized electron-ion nebula that is inflated behind the SN ejecta of a young magnetar. This model provides predictions for the radio synchrotron emission from the nebula for various input parameters such as the total magnetic energy, the nebula velocity, and the rate of energy input into the nebula.  Here we refer to models A, B, and C from \citet{Margalit2018b}, which correspond to unique model parameters and inferred source ages of $t_{\rm age,FRB121102}\approx 12$, $38$, and $13$ years, respectively. We refer the reader to the original paper for details of the models. 

In Figure~\ref{fig:frbmodels} we plot the SEDs and 9 GHz light curves for the three models, compared to the R20 sources. We find that the observed luminosities and minimum ages are consistent with the models for \repeater{}, especially if these sources represent a somewhat older population. For example, in the SED panel of Figure~\ref{fig:frbmodels} we find that the luminosities of the R20 sources are broadly consistent with the models if their ages are $\sim (1-3)\times t_{\rm age,FRB121102}$, or equivalently $\sim 30-100$ years. Similarly, in the light curve panel of Figure~\ref{fig:frbmodels} we find that the luminosities and timescales are consistent with models A and C (with $t_{\rm age,FRB121102}\approx 12$ and $13$ yr, respectively) but for older ages, while for model B the minimum ages are generally consistent with $t_{\rm age,FRB121102}\approx 38$ yr. Finally, we note that for older sources, the models predict that the flux will decay more slowly over the observed available baselines; thus, the observed shallow decays for some of the R20 sources are broadly consistent with the interpretation of $\gtrsim 30-100$ yr old sources.

We therefore conclude that a magnetar nebula model is thus far consistent with the luminosities and minimum age constraints for the R20 radio sources. Still, a better determination of their SEDs and light curve evolution are required to make a firm connection to this model. 

Finally, we note the possibility that the \repeater{} persistent radio source is due to an AGN. While this would weaken a distinction between FRB environments and AGN in dwarf galaxies, this does not discount the possibility that some of the R20 compact radio sources are LGRB afterglows.

\section{Occurence Rates}
\label{sec:rates}

We briefly compare the rate of compact radio sources in dwarf galaxies based on the sample of R20 to the FRB source density. In R20, the parent sample included 43,707 dwarf galaxies. Their observational efficiency was 0.75, since 37 sources in their initial target selection of 148 were not observed.  Thus, given 20 compact radio sources detected, their detection rate in the parent sample is $\sim 5\times 10^{-4}$.  

From the analysis of \citet{Nicholl2017c}, the estimated comoving volume density of repeating FRB-producing magnetar sources is $R_{\rm FRB}\tau = 10^4 \ \rm Gpc^{-3}$ (for a mean lifetime, $\tau$, and assuming that 10\% are beamed in our direction and that the active duty cycle is 30\%). Given a dwarf galaxy number density of $\approx 3\times 10^7 \ \rm Gpc^{-3}$ \citep{Faber2007}, the occurrence rate is $\sim 3\times 10^{-4}$.  Thus, the occurrence rate of the R20 sources is comparable to that of repeating FRBs.  There are clearly uncertainties in both estimates of at least a factor of few, including the fact that not all FRBs have been shown to reside in dwarf galaxies \citep{Bannister2019,Prochaska2019,Ravi2019,Marcote2020}, but the broad consistency indicates that the scenario proposed here is viable.  Furthermore, the occurrence of 1 in 5 FRBs in a dwarf galaxy is nevertheless higher than the fraction of star formation that takes place in dwarfs.

We further make the comparison to the expectation for LGRBs given the observed agreement between the afterglow models and the luminosities of the R20 sources (\S\ref{sec:lgrbs}). We use the observed rate from \citet{Wanderman2010}, $R_{\rm LGRB}(z)=1.3 (1+z)^{2.1} \ \rm Gpc^{-3} \ yr^{-1}$, integrated to $z=0.055$, and with a beaming factor of $50$ to find an expected all-sky rate of $\approx 4$ yr$^{-1}$. In the specific comparison to the R20 sample, the mean minimum age is $25$ years and the sky coverage is about $1/4$, leading to an expected $\sim 25$ LGRBs, which is in good agreement with the 20 observed sources and the R20 observational efficiency of 0.75.

\section{Conclusions and Future Observations}
\label{sec:conclusions}

We compared the properties of 20 compact radio sources located in dwarf galaxies and offset from the host centers to the properties of the persistent radio sources associated with \repeater{}, a few well-localized FRBs with non-detected persistent radio sources, PTF10hgi, and J1419+3940.  R20 argued that these sources are inconsistent with a star formation origin, normal SN remnants, and normal radio SNe, and instead conclude that they represent accreting massive black holes that have wandered from their galaxy centers.  Here we find that:

\begin{itemize}
    \item The radio luminosities of the R20 sources ($\sim 5\times 10^{26} - 10^{29} \ \rm erg \ s^{-1} \ Hz^{-1}$) are  comparable to the \repeater{} persistent radio source and our other comparison sources. Their generally lower luminosities compared to \repeater{} may be attributed to their older ages. 
    
    \item The majority of the R20 sources exhibit optically thin spectra between 1 and 9 GHz (a mean spectral index of $\alpha\approx -0.7$), consistent with the observed SED of the \repeater{} persistent source, as well as the late-time spectra of LGRB afterglows. Four of the 20 sources exhibit flat spectra with $\alpha\approx 0$.  
    
    \item The available crude light curves exhibit a range of behavior, including a $\sim 20 - 40\%$ decline for 6 sources over a temporal baseline spanning $4-18$ yr, a $\sim 20-90\%$ increase for 4 sources over $2-15$ yr, and flat or constant evolution for 3 sources over a period of $\sim 5-14$ yr. The remaining sources lack sufficient data for characterizing the temporal flux evolution. Given the relatively recent discovery of \repeater, the long-term evolution of its persistent radio emission has not yet been characterized; however, if these sources are older than \repeater, we expect them to evolve over much longer timescales.
    
    \item The distribution of spatial offsets for the R20 sources is consistent with the observed offsets for \repeater{}, the other well-localized FRBs, PTF10hgi, and J1419+3940. Similarly, the range of offsets is largely consistent with the observed distributions for LGRBs and SLSNe. 
    
    \item The R20 sources span radio-to-optical flux ratios of $R\approx 0 - 2.2$. While some of the sources are therefore consistent with the observed ratio of $2$ for \repeater{}, the lower ratios in the other cases may be indicative of a somewhat older population relative to \repeater{}. This is further borne out by the consistency with the range of ratios for J1419+3940 over the 22 year duration of its light curve.
    
    \item The luminosities and minimum ages of the R20 sources are consistent with the expectation for off-axis LGRBs, as well as the light curve of J1419+3940, which has been argued to be powered by an orphan LGRB afterglow or a magnetar nebula. 
    
    \item The luminosities and minimum ages of the R20 sources are consistent with a model of a magnetized electron-ion nebula invoked to explain \repeater. The generally lower luminosities relative to \repeater{} may point to somewhat older ages of $\sim 30-100$ years.
    
    \item The occurrence rate of compact offset radio sources in dwarf galaxies implied by the R20 sample ($\sim 5\times 10^{-4}$) is consistent with the inferred repeating FRB source density as well as the number of expected LGRB off-axis afterglows.
\end{itemize}

We thus find that based on the available data the R20 sources are consistent with sharing a common origin with \repeater{}, and perhaps PTF10hgi and J1419+3940.  To test this association more robustly, we require a simultaneous determination of the radio SEDs and time evolution.  Within the magnetar nebula framework we expect a correlation between the SED shape and time evolution (Figure~\ref{fig:frbmodels}).  In addition, continued monitoring will better constrain the source variability, particularly in comparison to AGN variability; this can be probed both in the radio (e.g., \citealt{Valtaoja1992,Turler2000,Lindfors2006}) as well as in the optical regime, where the long-term optical variability has been successfully used to identify AGN in low-mass galaxies (\citealt{Baldassare2018}; although we note that the bulk of these galaxies are spectroscopically identified as AGN, unlike the R20 sources discussed here).

As discussed in R20, X-ray detections would help to confirm the AGN nature of these sources \citep{Baldassare2017}, but even non-detections cannot definitively rule out the presence of low-luminosity, radio-loud AGN \citep{Mauch2007}. Indeed, five such sources were discovered on the basis of their radio emission \citep{Park2016}, although in more massive galaxies ($\sim 10^{10} M_{\odot}$) than those in the R20 sample.

Finally, VLBI observations can be used to further constrain the physical scale of these sources and to search for scintillation-induced variability, which is only expected in the magnetar nebulae scenario. Similarly, high resolution optical imaging with the Hubble Space Telescope will allow for a more careful assessment of the host galaxy properties at the locations of the radio sources, particularly in regard to regions of high star formation, which would be indicative of magnetar or LGRB progenitors \citep{Fruchter2006}.

\acknowledgments \textit{Acknowledgments.} We thank the referee for their careful assessment of this work. We thank Amy Reines and Jenny Greene for providing data from the R20 paper as well as for helpful discussions regarding the original work. We also thank Huib Intema for providing information about the GMRT 150 MHz All-sky Radio Survey data for J0134$-$0741. We thank Kate Alexander, Yvette Cendes, Sebastian Gomez, and Matt Nicholl for helpful discussions. The Berger Time-Domain Group is supported in part by NSF grant AST-1714498. BM is supported by NASA through the NASA Hubble Fellowship grant \#HST-HF2-51412.001-A awarded by the Space Telescope Science Institute, which is operated by the Association of Universities for Research in Astronomy, Inc., for NASA, under contract NAS5-26555. 

\software{Astropy \citep{astropy2018}, \texttt{BOXFIT} \citep{vaneerten2011}, CASA \citep{McMullin2007}, \texttt{extinction} \citep{Barbary16}, pwkit \citep{Williams2017}}

\begin{deluxetable*}{lccccccc}
\small
\caption{Compact Radio Sources from R20}
\tablehead{
\colhead{Source} & 
\colhead{RA} & 
\colhead{Dec} & 
\colhead{$S_{\rm NVSS}$\footnote{1.4 GHz flux density from NVSS.}} &
\colhead{$S_{\rm VLASS}$}\footnote{3 GHz flux density from VLASS.} & 
\colhead{$t_{\rm min}$}\footnote{Minimum age of each source based on a detection in either NVSS or FIRST.} &
\colhead{$\alpha$\footnote{Spectral index ($F_\nu \propto \nu^\alpha$) between 1.4 and 9 GHz (see Figure~\ref{fig:seds}).}} & 
\colhead{$R$\footnote{Radio-to-optical flux ratio defined as $R=\rm{log}(F_{\rm{1.4 GHz}}/S_r)$}} \\ 
\colhead{} & 
\colhead{[J200]} & 
\colhead{[J200]} & 
\colhead{[mJy]} &
\colhead{[mJy]} &
\colhead{[yr]} &
\colhead{}& 
\colhead{}
}  
\startdata
J0019+1507 & 00:18:59.9856 & +15:07:11.028 & 3.4 &2.1 & 25.5 & 0.02 & 0.90\\
J0049-0242	& 00:49:51.8448	& -02:42:43.020 & <$2.5^{\dagger}$ & 1.5 & 8.6 & -0.56 & 0.32\\
J0106+0046	& 01:06:07.3080	& +00:46:34.320 & 1.79 & 1.1 & 25.9 & -0.79 & 0.06\\
J0134-0741	& 01:34:08.7408	& -07:41:45.348 & 17.1 & 22.2 & 25.5 & 0.03 & 1.32\\
J0247+0535	& 02:47:47.3256	& +05:35:15.576 & 3.0 & 1.8 & 25.5 & -0.38 & 1.13\\
J0854-0240a	& 08:54:31.7472	& -02:41:00.060 & <$2.5^{\dagger}$ & 1.1 & 21.1 & -0.81 & 0.81\\
J0854-0240b	& 08:54:31.8336	& -02:40:58.944 & <$2.5^{\dagger}$ & 0.5 & 21.1 & -0.78 & 0.51\\
J0903+4824	& 09:03:12.9672	& +48:24:13.716 & 8.4 & 2.7 & 25.5 & -1.55 & 1.30\\
J0906+5610	& 09:06:13.7688	& +56:10:15.132 & 4.7 & 2.3 & 25.5 & -0.87 & 1.07\\
J0909+5655	& 09:09:08.6904	& +56:55:19.740 & <2.5 & 0.6 & 22.4 & -0.58 & 0.98\\
J0931+5633	& 09:31:38.4192	& +56:33:19.872 & 8.9 & 4.7 & 25.2 & -0.85 & 2.29\\
J1027+0112	& 10:27:41.3784	& +01:12:06.444 & 4.4 & 3.3 & 25.5 & -0.20 & 1.51\\
J1136+1252	& 11:36:48.5256	& +12:52:39.900 & <2.5 & 2.3 & 19.8 & 0.08 & 0.71\\
J1136+2643	& 11:36:42.5760	& +26:43:35.652 & 1.3 & 1.8 & 24.9 & -0.72 & 0.68\\
J1143-0325	& 11:43:18.6840	& -03:25:52.644 & 6.0 & 2.7 & 25.5 & -0.83 & 1.40\\
J1200-0341	& 12:00:58.3008	& -03:41:18.456 & 4.1 & 3.0 & 25.5 & -0.55 & 0.67\\
J1220+3020	& 12:20:11.2656	& +30:20:08.304 & <2.5 & 0.8 & 26.5 & -0.50 & 0.18\\
J1226+0815	& 12:26:03.6432	& +08:15:19.008 & 2.9 & 1.6 & 25.5 & -0.95 & 0.78\\
J1253-0312	& 12:53:05.9688	& -03:12:58.752 & 5.7 & 3.1 & 25.5 & -0.70 & 0.34\\
J2359+1435	& 23:59:45.1464	& +14:35:02.184 & <2.5 & 8.8 & 8.5 & -0.05 & 0.03\\
\enddata
\tablecomments{$\dagger$ Cannot derive accurate flux densities for these sources due to confusion with nearby bright sources.}
\label{tab:vla}
\end{deluxetable*}

\begin{deluxetable*}{lcccc}
\small
\tablecolumns{5}
\caption{Archival Radio Observations}
\tablehead{
\colhead{Source} & 
\colhead{Frequency} &
\colhead{Flux} &
\colhead{Date} & 
\colhead{Program} \\ 
\colhead{} & 
\colhead{[GHz]} & 
\colhead{[mJy]} & 
\colhead{[UT]} & 
\colhead{}
}  
\startdata
J0106+0046  & 1.5 &1.93 $\pm$ 0.10& 2008 Sep&Stripe 82 VLA (13B-372) \\
J0134$-$0741 & 0.15 & 46.0 $\pm$ 8.1& 2011 Nov 1 & GMRT 150 MHz All-Sky Survey\\
J1136+2643 & 1.5 & <0.64 & 1984 May 25 & VLA AJ0108 \\
J1200$-$0341  & 1.4 &7.9 $\pm$ 1.3 & 2003 Feb 16 &VLA AG0644\\
J1220+3020 & 1.5 & <0.68 & 1984 May 29 & VLA AJ0108 \\
...& 1.5 & <0.76 & 1988 Jul 12 & VLA AB0506 \\
\enddata
\tablecomments{Limits correspond to $3\sigma$.}
\label{tab:archival}
\end{deluxetable*}

\bibliography{ref}
\bibliographystyle{apj}

\end{document}